# Quis pendit ipsa pretia: facebook valuation and diagnostic of a bubble based on nonlinear demographic dynamics

Peter Cauwels and Didier Sornette

**Abstract:** We present a novel methodology to determine the fundamental value of firms in the social-networking sector, motivated by recent realized IPOs and by reports that suggest sky-high valuations of firms such as facebook, Groupon, LinkedIn Corp., Pandora Media Inc, Twitter, Zynga. Our valuation of these firms is based on two ingredients: (i) revenues and profits of a social-networking firm are inherently linked to its user basis through a direct channel that has no equivalent in other sectors; (ii) the growth of the number of users can be calibrated with standard logistic growth models and allows for reliable extrapolations of the size of the business at long time horizons. We illustrate the methodology with a detailed analysis of facebook, one of the biggest of the social-media giants. There is a clear signature of a change of regime that occurred in 2010 on the growth of the number of users, from a pure exponential behavior (a paradigm for unlimited growth) to a logistic function with asymptotic plateau (a paradigm for growth in competition). We consider three different scenarios, a base case, a high growth and an extreme growth scenario. Using a discount factor of 5%, a profit margin of 29% and 3.5 USD of revenues per user per year yields a value of facebook of 15.3 billion USD in the base case scenario, 20.2 billion USD in the high growth scenario and 32.9 billion USD in the extreme growth scenario. According to our methodology, this would imply that facebook would need to increase its profit per user before the IPO by a factor of 3 to 6 in the base case scenario, 2.5 to 5 in the high growth scenario and 1.5 to 3 in the extreme growth scenario in order to meet the current, widespread, high expectations. To prove the wider applicability of our methodology, the analysis is repeated on Groupon, the well-known deal-of-the-day website which is expected to go public in November 2011. The results are in line with the facebook analysis. Customer growth will plateau. By not taking this fundamental property of the growth process into consideration, estimates of its IPO are wildly overpriced.

Peter Cauwels, Department of Management, Technology and Economics, ETH Zurich, Kreuzplatz 5, 8032 Zurich, Switzerland, pcauwels@ethz.ch

Didier Sornette, Department of Management, Technology and Economics, ETH Zurich, Kreuzplatz 5, 8032 Zurich, Switzerland and Swiss Finance Institute c/o University of Geneva, 40 Blvd Du Pont D'Arve, 1211 Geneva 4, Switzerland, dsornette@ethz.ch

On September 7, Reuters announced that facebook's revenues for the first half of 2011 had doubled to 1.6 billion USD. Its net income over that same period came to almost 500 million USD (Oreskovic [2011]). According to facebook's website (facebook [21 January 2011]), the company raised 1.5 billion USD from investors including Goldman Sachs and Digital Sky Technologies last January 2011. This was done at a valuation of roughly 50 billion USD.

Rumors, picked up by popular business media like the Financial Times or CNBC, anticipate facebook to go public in 2012. The expectations for its IPO are high. Sketchy valuations range from "more than 66.5 billion USD" (Dembosky [2011]) and "roughly 80 billion USD" (Oreskovic [2011]) up to "north of 100 billion USD" (Kelly [2011]). No specifics are given in any of these references on how these valuations were obtained and which methodology was used.

These hazy figures, however, would value facebook at a price earnings ratio between 50 and 100. For the more critical investor, such numbers resonate memories from a not-so-distant past when, at the turn of the millennium, at the height of the Internet bubble, the average price earnings ratio for the S&P500 peaked at an irrationally exuberant 40 mark (Shiller [2005]). At that time, one of the most important rationalizations for the sky-high valuation of dotcom firms was based on the "real option" pricing of the new universe of possibilities offered by the Internet (see Mauboussin and Hiler [1999]): the higher the uncertainty and variability of future potential opportunities, the higher the price of the real option supposed to be embedded in dotcom firm stocks. In the same spirit, pundits have argued that "Even at 600-plus million users and several billion dollars in annual revenue, facebook is small, relative to the opportunity going forward" (quote of Kevin Werbach, a Wharton legal studies and business ethics professor, in Knowledge@Wharton (25 May 2011)). Many also see facebook as having strong potential for further growth, and this is taken to justify the large implied valuation mentioned above. In the sequel, we provide quantitative evidence against this view.

Groupon, the well-known deal-of-the-day website, which is expected to go public in November 2011, has been subject to similar media speculation. In June 2011, after announcing its upcoming IPO, the company was supposedly worth "roughly 30 billion dollars" (see e.g. Rusli and De La Merced [2011]). In the most recent amendment to its Form S1 filing to the SEC, the company's expectations had cooled down to a more moderate 11 billion dollars valuation (see SEC [2011]), corresponding to a share price of 17 dollars.

In line with the roman poet Decimus Junius Juvenalis, who questioned who would be watching the watchmen themselves, we ask a similar question: Quis pendit ipsa pretia? Who values these actual valuations? Investment sentiment is currently at a historical low and market confidence is shattered by a banking crisis that has turned into a government debt crisis. It should be carefully considered what the impact might be for the stability of financial markets and the global economy if a new bubble were built on top of these fragile bases.

We propose a new methodology that makes a robust valuation of companies such as facebook possible. The novel valuation method builds on the rather unique characteristics of social-networking companies, namely that they are all about advertising and services that consider the user (and its personal information) as the "product" being tracked and sold. Therefore, their revenues and profit, and as such their value, are inherently linked to their user (or customer) basis. When analyzing the historical evolution of the number of users, which is periodically published on the facebook press website, it can be seen that the population growth recently underwent a change in regime. Under pressure of competition, a limited amount of user devices, impenetrable markets and, a fortiori, a limited world population, the growth of facebook users has clearly started deviating from the exponential function (a paradigm for unlimited growth) and is perfectly following the track of the logistic function (a paradigm for growth in competition with saturation). We calibrate three logistic growth functions, each corresponding to a different scenario: a base case, a high growth and an extreme growth scenario. Using a discount factor of 5%, a profit margin of 29% and 3.5 USD of revenues per user per year yields a value of facebook of 15.3 billion USD in the base case scenario, 20.2 billion USD in the high growth scenario and 32.9 billion USD in the extreme growth scenario. According to our model, this would imply that facebook would need to increase its profit per user before the IPO by a factor of 3 to 6 in the base case scenario, 2.5 to 5 in high growth scenario and 1.5 to 3 in the extreme growth scenario in order to meet the current, widespread, high expectations.

The organization of the paper is as follows. Section 1 describes the available data on the growth of the number of facebook users. Section 2 presents evidence for a transition from an early purely exponential growth to a growth in competition. Section 3 introduces the logistic growth model and presents its calibration to the data on the number of facebook users. Three scenarios for the future number of facebook users are obtained. Section 4 converts these three scenarios into a valuation of facebook, using conservative assumptions to avoid devaluing the company unnecessarily. To prove the wider applicability of our methodology, the analysis is repeated on Groupon, the well-known deal-of-the-day website which is expected to go public in November 2011. This is done in Section 5. Finally, Section 6 summarizes and concludes.

## Facebook's users

Facebook is a social network where users build and maintain their own personal profile, connect to each other and exchange personal information. This usually hard-to-get private information is the raw material used by facebook to customize advertisement. The majority of facebook's revenues, hence profit, is generated by advertisement. This is the basis for a smart and robust business plan. If a user puts in her profile that she lives in Zurich and loves skiing and hiking, she may receive advertisement from an outdoor shop nearby. It is clear that facebook's core asset is its user basis; if its users grow, its revenues and profit will grow. Whereas most other companies need to pay for detailed proprietary customer information, facebook receives this for free due to its unique core application (Knowledge@Wharton, 23 January 2008, [25 May 2011]).

The number of users is published regularly on facebook's website; Exhibit 1 gives the results (facebook, http://www.facebook.com/press/info.php?statistics and facebook [2011]).

**Exhibit 1: Number of facebook users**

| Publication Date | Users (million) |
|---|---|
| July 2011 | 750 |
| July 2010 | 500 |
| February 2010 | 400 |
| December 2009 | 350 |
| September 2009 | 300 |
| July 09 | 250 |
| April 2009 | 200 |
| February 2009 | 175 |
| January 2009 | 150 |
| August 2008 | 100 |
| October 07 | 50 |
| April 07 | 20 |
| December 06 | 12 |
| December 05 | 5.5 |
| December 04 | 1.0 |

*Note: Taken from ref. facebook [2011]*

In this study, the future growth of users will be regarded as the key to the future valuation of the company. The methodology goes as follows. Firstly, a growth model will be proposed that allows forecasting facebook's future user basis. Three scenarios will be proposed: a base case, a high growth and an extreme growth scenario. Facebook is valued under each of these scenarios, based on a one-dollar profit per user per year. This will be done for a range of discount factors. In this way, hard numbers are separated from soft numbers. Hard numbers are considered to be the number of users and their forecast under the different scenarios. Soft numbers are the dollar profit per user and the discount factor. Finally, a cautious proposal will be done on which soft numbers may be used. This will result in our valuation of facebook for the three different scenarios. The reader may, however, apply any other soft numbers at his or her own discretion using our forecasts of the hard numbers.

**Exponential growth versus growth in competition**

**Pure exponential growth regime**

When assessing the future growth of a population, a new technology, complexity in the universe, a company or more specifically the number of users of a social medium, it is important to make a clear distinction between unlimited growth and growth in competition (see e.g. Modis [2002], [2003] and [2009]). One example of unlimited growth is proportional or exponential growth. This is growth without any boundary conditions. Suppose the number of users of facebook and hence the revenues and profits could rise without any limitation of resources and without any battle lost against competitors. What would the growth process of the user basis and as such the growth of the revenues and profit look like?

**Exhibit 2**
**The evolution of facebook users (semi-log representation)**

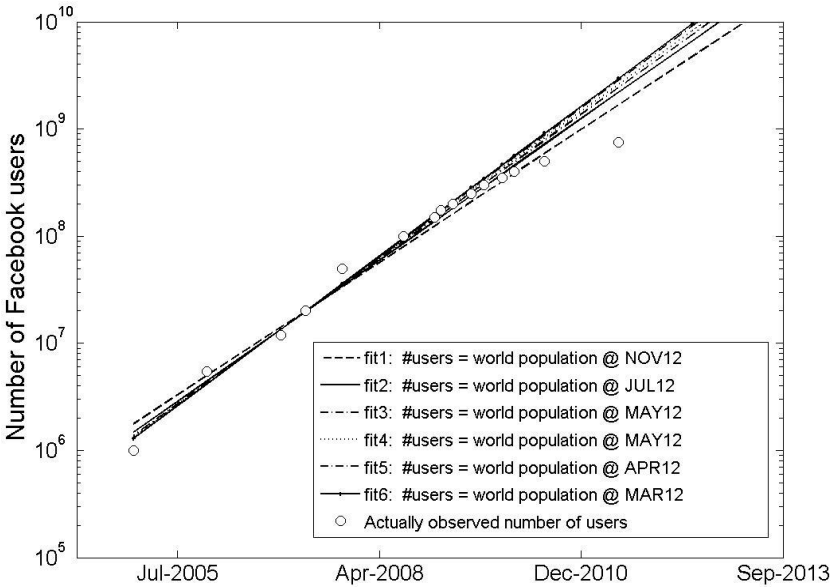

*Note: Fit 1 is done on all data-points; fit 2 to fit 6 are the results of recursively omitting the last data point, the last two data points, until the last 6 data points.*

Exhibit 2 shows the future number of facebook users when an exponential extrapolation is used. There are 6 different models. Fit 1 is done on all data-points; fit 2 to fit 6 are the results of recursively omitting the last data point, the last two data points, until the last 6 data points. From this figure, it can be seen that, when using an exponential extrapolation, the number of facebook users surpasses, in any scenario, the world population (of 7 billion, Bloom [2011]) before the end of 2012. As the world population is a hard constraint to the number of facebook users, it is clear that an unlimited growth process is not suitable in this context.

**Growth in competition and logistic function**

When one takes a closer look at Exhibit 2, more specifically at the last two user counts, it can be seen that the growth of the number of users has started leveling off. This is the result of a growth process under competition, with boundary conditions (like the world population), or, limitations to growth. In such a situation, the natural growth law is the S-shaped logistic curve or S-curve. This pattern is characteristic of a species population growing under Darwinian competition such as a pair of rabbits on a fenced-off range. There is a population explosion in the beginning. However, the available food can feed only a limited amount of rabbits. As the population approaches its limit, the growth rate slows down. Eventually, as explained in Griliches [1988] and Modis [2002], the population stabilizes as the S-curve reaches its ceiling.

Theodore Modis [2009] describes this as follows: "*Whenever there is growth in competition (survival of the fittest), a "population" will evolve along an S-curve, be it sales of a newly launched product, the diffusion of a new technology or idea, an athlete's performance, or the life-long achievement of an artist's creativity. And because every niche in nature - and in the marketplace - generally becomes filled to completion, S-curves possess predictability.*"

The same paradigm applies to the number of facebook users. In the beginning, there is exponential growth. However, under the constraint of a limited amount of user devices (such as smart-phones or pc's) and a fortiori a limited world population (of 7 billion, Bloom [2011]), the growth rate decays, the number of users stabilizes and will reach a ceiling. As such, its idiosyncratic growth process stops and further growth is limited to systemic growth, like general growth of the world population, global GDP or of similar technological constraints. As can be seen in Exhibit 2, the most recent user counts are evidence of this imminent ceiling.

Let us demonstrate this fact in a more quantitative manner. An exponential function is fitted to the data points from Exhibit 1, using an ordinary least squares method, based on the fitting error given in equation 1:

$$fitting\ error = \frac{1}{df} \sum \frac{(o_i - e_i)^2}{e_i^2} \qquad (1)$$

In this equation, df is the number of data points, $o_i$ the observations and $e_i$ the fitted function. The last observation is recursively omitted from the analysis. This is done 10 times. So, the point 0 on the x-axis gives the result for a fit on the full data set, whereas the point 10 gives the result after excluding the 10 most recent observations, i.e., using data up to October 2007 when the number of facebook users was 50 million. Exhibit 3 plots the fitting error, defined in equation 1, as a function of the number of omitted data points. This approach corresponds to plotting the fitting error as a function of time going backward.

**Exhibit 3**
**The evolution of the fitting error (equation 1) of an exponential function when the most recent data points are recursively omitted from the data set (semi-log representation)**

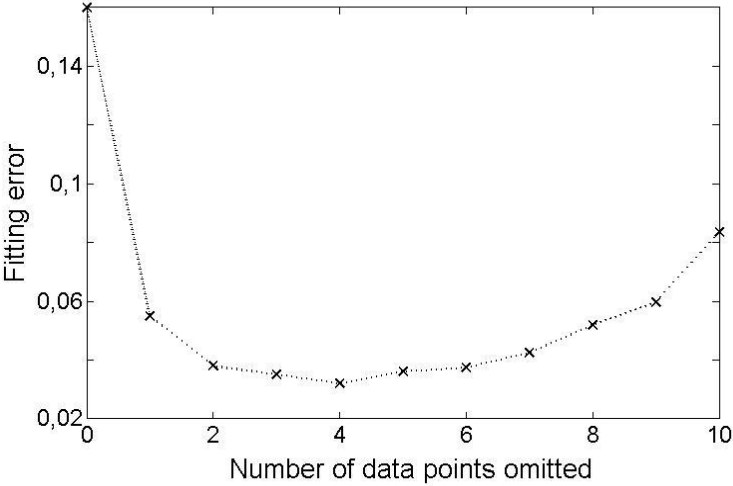

The figure reveals a distinct pattern. Going from right to left, the fitting error decreases gradually as the number of data points increases and plateaus for included data up to February 2010 when the number of users was 400 million. The existence of a well-defined plateau of the fitting error qualifies the exponential model for the growth of the number of facebook users until February 2010. However, adding the last two data points changes the picture completely; the fitting error jumps up a factor of 3, from 0.055 up to 0.16. A change of regime appears clearly in the data set, showing that the pure exponential growth is not longer the correct mechanism to describe the growth of the facebook user base. This is first evidence of a stabilization of the number of users. It demonstrates that the number of facebook users follows a growth process under competition and proves that a more suitable model is needed than an exponential growth model to provide a quantitative forecast.

In the next section, we will explain the properties of the logistic function that accounts for the growth under competition and explain the observed change of regime illustrated in Exhibit 3. We will introduce a fitting procedure that will result in an efficient forecasting. This will make it possible to design different growth scenarios and eventually to bracket the fundamental value of facebook.

## The logistic growth model and its calibration

### The logistic equation

Let P represent a population size, r the initial growth rate and K the carrying capacity. Then, the growth of a population in competition can be described by the following "logistic" differential equation:

$$\frac{dP}{dt} = rP\left[1 - \frac{P}{K}\right] \qquad (2)$$

In this equation, when the population is much smaller than the carrying capacity (P<<K), its growth is exponential (being proportional) with a rate r. At the other extreme, when the population reaches the carrying capacity (P=K), the growth stops and the population remains constant. It has reached a ceiling, which is the carrying capacity (K), by definition.

The complete solution to this differential equation, with $P_0$ being the initial population ($P(t=0)=P_0$), is called the logistic function. It can be written as follows:

$$P(t) = \frac{K P_0 e^{rt}}{K + P_0(e^{rt} - 1)} \qquad (3)$$

This is the function that will be fitted to our user data. A stepwise procedure is proposed. First, r, K and $P_0$ are estimated using a straightforward analytical approach. These analytical estimates will be used as input values for a subsequent least squares optimization in the three dimensional space of r, K and $P_0$. The final result will be our base case scenario for future user growth. Next, the one-sided 80% and 95% confidence intervals of K are calculated. This is the direct result of the previous analytical parameter estimation using t-distribution statistics. Two additional least squares optimizations are done. This time, K is fixed either at its 80% or 95% confidence value and the optimization is done in the two dimensional space of r and $P_0$. This will provide the parameters for our high growth and extreme growth scenarios. In the following, we will explain step by step how this is done.

### Calibration using growth rates

Let us call the counts in Exhibit 1 $P_i$ at each time $t_i$. Then, the discrete growth rate $R^d_i$ between two observations $P_i$ and $P_{i-1}$ at time $t_i$ and $t_{i-1}$ is calculated as follows:

$$R^d_i = \frac{\ln\left(\frac{P_i}{P_{i-1}}\right)}{(t_i - t_{i-1})} \qquad (4)$$

On the other hand, the continuous growth rate $R^c_i$ of the logistic function can be directly derived from equation 2:

$$R^c_i = \frac{1}{P}\frac{dP}{dt} = r\left[1 - \frac{P}{K}\right] \qquad (5)$$

The discrete observations of the growth rate $R^d_i$ can be fitted to the continuous growth rate function $R^c_i$ of the logistic function according to:

$$\frac{\ln\left(\frac{P_i}{P_{i-1}}\right)}{(t_i - t_{i-1})} = r\left[1 - \frac{\frac{P_i + P_{i-1}}{2}}{K}\right] \qquad (6)$$

In other words, we express $y=R^d_i$ as a function of $x=P_i+P_{i-1}$ using the linear regression $y=a+bx$, where the regressions coefficients $a$ and $b$ allow us to determine $r=a$ (the initial growth rate) and $K = -a/2b$ (the carrying capacity, or the maximum number of future facebook users). The linear regression, shown in Exhibit 4, gives a=1.40 and b=-1.74 $10^{-9}$. Additionally, using t-distribution statistics, the confidence intervals for K are estimated. Exhibit 4 provides a preliminary support for the validity of the logistic growth model (2) and (3).

**Exhibit 4**
**The discrete growth rate is a linear function (y = -1.74 $10^{-09}$ x + 1.40) of the population (equation 6). This gives the initial growth rate, r, and the carrying capacity, K**

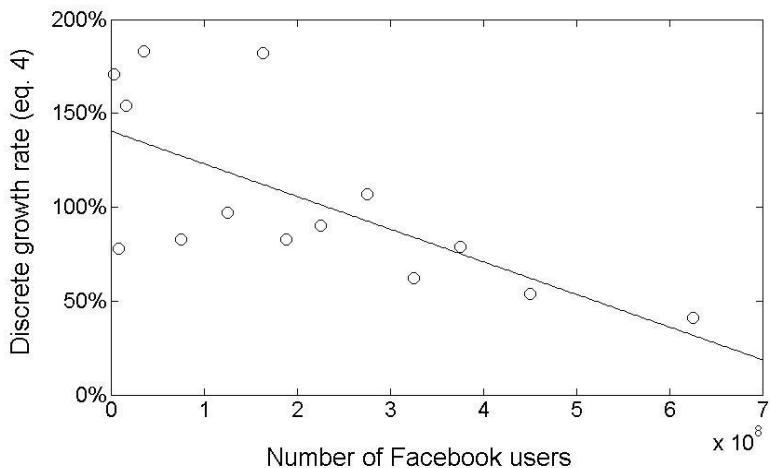

Once the values for r and K are known, the only remaining unknown parameter of the logistic function is $P_0$, which is the initial population. By rearranging terms in equation 3, $P_0$ can be calculated for every observation $P_i$ at time $t_i$ (in Exhibit 1) as:

$$P_{0,i} = \frac{-P_i K}{P_i(e^{r_i t}-1) - K e^{r_i}} \qquad (7)$$

The results for r and K from the linear regression obtained from equation 6 and the average of the calculated $P_{0,i}$ in equation 7 will be used as input values for a subsequent least squares optimization. Exhibit 5 gives an overview.

**Exhibit 5**
**Results from the preliminary linear regression of the average growth rate versus facebook user population size**

|  | **Result** |
|---|---|
| **K (avg.)** | 0.81 billion |
| **K (80%)** | 1.11 billion |
| **K (95%)** | 1.82 billion |
| **r** | 1.40 / year |
| **$P_0$** | 234 thousand |

**Full calibration of the logistic equation and the three growth scenarios**

We complement this preliminary analysis by three different least squares minimizations of the fitting error defined in equation 1:

- An optimization in the three dimensional space of r, K and $P_0$ using K (avg.), r and $P_0$ from Exhibit 5 as input values. This result will be called our base case scenario;
- An optimization in the two dimensional space of r and $P_0$ keeping K (80%) fixed and using r and $P_0$ from Exhibit 5 as input values. This result will be called our high growth scenario;
- An optimization in the two dimensional space of r and $P_0$ keeping K (95%) fixed and using r and $P_0$ from Exhibit 5 as input values. This result will be called our extreme growth scenario;

The results, together with the fitting error (equation 1), are given in Exhibits 6 and 7.

**Exhibit 6**
**The parameters that will be used for the three different scenarios**

|  | **Scenario** | | |
|---|---|---|---|
|  | **Base case** | **High growth** | **Extreme growth** |
| **K (billion)** | 0.84 | 1.11 | 1.82 |
| **$P_0$ (thousand)** | 423 | 597 | 647 |
| **r (per annum)** | 1.26 | 1.16 | 1.12 |
| **Fitting error** | 0.020 | 0.024 | 0.037 |

**Exhibit 7**
**The three different facebook user growth scenarios fitted to the observations**

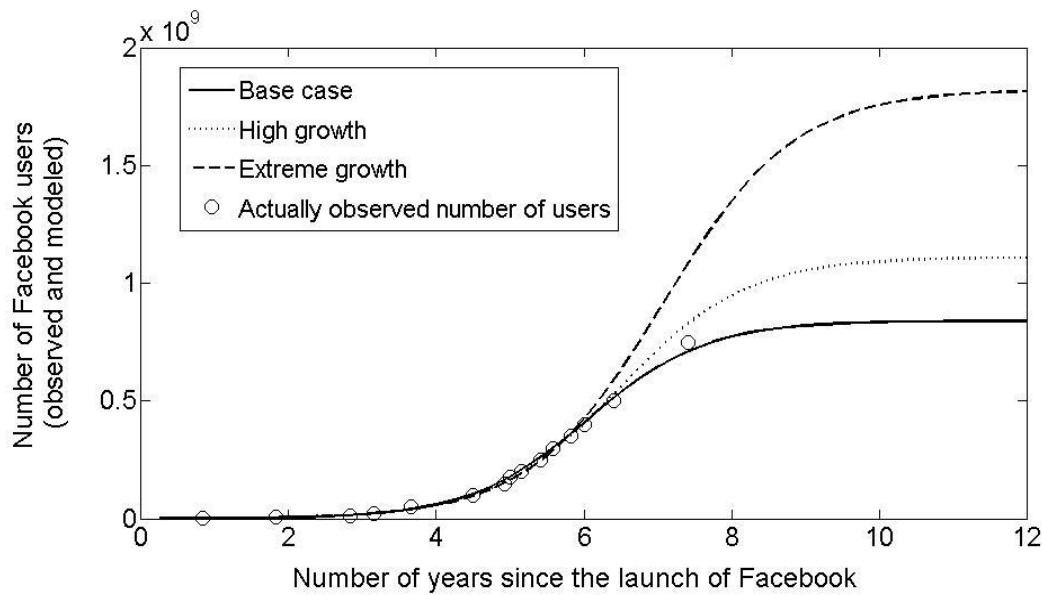

The fitting error may be compared with the results of the exponential fitting exercise shown in Exhibit 3. Compared with the 0.16 fitting error of the exponential function on the full dataset, all three scenarios give better fitting results, as should be expected from the fact that the exponential model is nested in the logistic model, which has one additional parameter. Comparing the base case logistic growth scenario with the pure exponential growth model, the later (null hypothesis) is strongly rejected with a *p*-value of less than 0.001. This means that the addition of the carrying capacity K is highly significant statistically and that the data contains already enough information to bracket its value.

In the next section, a valuation of facebook will be made in the three different scenarios. To prove the wider applicability of our methodology, the analysis is repeated on Groupon, the well-known deal-of-the-day website in the subsequent section.

## The valuation of facebook
**Normalized valuation given a fixed profit of one dollar per user per year**

Facebook is a private company; it does not publicly disclose its financial statements. Nevertheless, so-called "sources with knowledge of its financials" tend to leak out results on a regular basis. As we do not have access to official figures, we will, in a first step, calculate a normalized valuation. This is the value of the company given a fixed profit of one dollar per user per year. Next, unofficial revenue and profit estimates as reported by different business media will be used as soft numbers to give a personal best-estimate valuation of the company. Thanks to this approach, the reader will be able to calculate any alternative valuation using personal soft numbers at his or her own discretion.

The normalized valuation of facebook is straightforward to calculate once the future number of users is known. Let us assume that the company generates cash flows on a yearly basis for the next 50 years. Taking a one-dollar profit per user per year, each yearly cash flow equals the amount of forecasted users in that year. We conservatively assume that all the profit is distributed to the shareholders. By discounting these cash flows and adding all the present values up, we come to the final normalized valuations. It should be mentioned that any future profit growth, due to the general growth of the global economy, is taken into account by choosing an adequate discount rate, such as a suitable benchmark rate (e.g. yield on ten year treasury notes) minus an estimation of future inflation.

Exhibit 8 gives the result of this exercise for the three different scenarios.

**Exhibit 8**
**The normalized valuation of facebook using different discount factors assuming a one-dollar profit per user per year over 50 years**

| facebook's value (in billion USD) for 1 USD profit per user per year | | | |
|---|---|---|---|
| Discount Factor | Base Case | High Growth | Extreme Growth |
| 2% | 26.4 | 34.8 | 56.9 |
| 3% | 21.6 | 28.4 | 46.5 |
| 4% | 18.0 | 23.7 | 38.8 |
| 5% | 15.3 | 20.2 | 32.9 |
| 6% | 13.2 | 17.4 | 28.4 |
| 7% | 11.6 | 15.2 | 24.8 |
| 8% | 10.2 | 13.5 | 21.9 |
| 9% | 9.2 | 12.1 | 19.6 |
| 10% | 8.3 | 10.9 | 17.7 |

**Valuation based on financial results circulating in the business media**

Let us now make a cautious estimate of what an acceptable yearly profit per user might be. Exhibit 9 summarizes the financial results that are circulating throughout the business media (Carlson, [2011], Tsotsis [2011], Wikipedia, [2011]). It is clear that facebook has delivered excellent operational results with an average profit margin of 29% over the last 3 years and a growth in revenues in line with its exponential user growth (until February 2010).

**Exhibit 9**
**Facebook's financial results circulating throughout the business media**

| Financial Year | Revenues (million USD) | Profit (million USD) | Profit Margin |
|---|---|---|---|
| 2006 | 52 | | |
| 2007 | 150 | | |
| 2008 | 280 | | |
| 2009 | 775 | 200 | 26% |
| 2010 | 2000 | 600 | 30% |
| 2011E | 3200 | 1000 | 31% |

**Exhibit 10**
**The evolution of the revenues (crosses; y=7.6e$^{0.84x}$) and the actual users (circles; y=7.5e$^{1.04x}$) of facebook since the launch of the company (semi-log representation)**

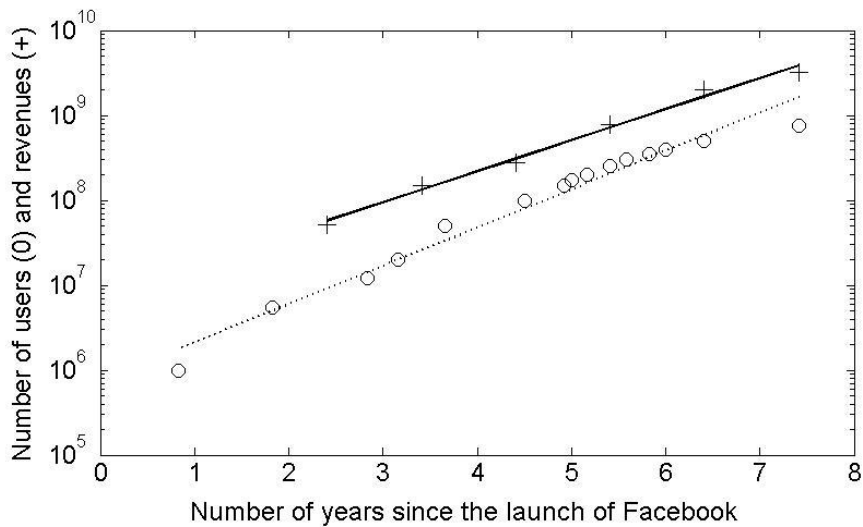

Exhibit 10 shows that the revenues and the user growth have both been growing approximately exponentially since the launch of the company (7.5 years ago). These results can be used to estimate the revenues from the number of users, $\Delta t$ being the number of years since the launch of the company.

$$\frac{Revenues}{Users} = \frac{7.6 \cdot 10^6 \, e^{0.84 \Delta t}}{0.75 \cdot 10^6 \, e^{1.04 \Delta t}} = 10 \, e^{-0.20 \Delta t} \tag{8}$$

These results shows that, on average, the revenues per year per user have halved every 3.5 years (this is calculated as ln(2)/0.20). Let us now take this revenue decay into consideration and calculate the average revenue per user over the last five years using eq. 8:

$$\left(\frac{Revenues}{Users}\right)_{Avg} = 10 \left(\frac{e^{-0.20*7.5} + e^{-0.20*6.5} + e^{-0.20*5.5} + e^{-0.20*4.5} + e^{-0.20*3.5}}{5}\right) = 3.5 \tag{9}$$

Thus, when we use the average profit margin of 29% (from Exhibit 9) and the revenue per user from equation 9, we arrive at an estimated profit of 1.0 USD per user per year. This is exactly the number that was used in the normalized valuation of the previous section.

Let us suppose, very conservatively, that the average benchmark yield equals the average inflation so that real interest rates remain at 0% over the next 50 years. In that case, the discount factor to be used in the valuation is equal to the equity risk premium. According to Fernandez [2011], who did a survey with 5731 answers, the equity risk premium used in 2011 for the USA by professors averaged 5.7%, by analysts 5.0% and by companies 5.6%.

Using a discount factor of 5%, a profit margin of 29% and 3.5 USD of revenues per user per year gives a value of facebook of 15.3 billion USD in the base case scenario, 20.2 billion USD in the high growth scenario and 32.9 billion USD in the extreme growth scenario.

## The valuation of Groupon

To prove that the proposed methodology can be easily applied to value other social networking companies we applied it to Groupon. The company is expected to go public in November 2011. The valuation that is proposed in the SEC filing form S1 is 17 dollar per share (see SEC [2011]). With about 630 million shares outstanding, this would correspond to a market capitalization of 10.7 billion USD.

**Exhibit 11**
**The three different Groupon customer growth scenarios fitted to the observations**

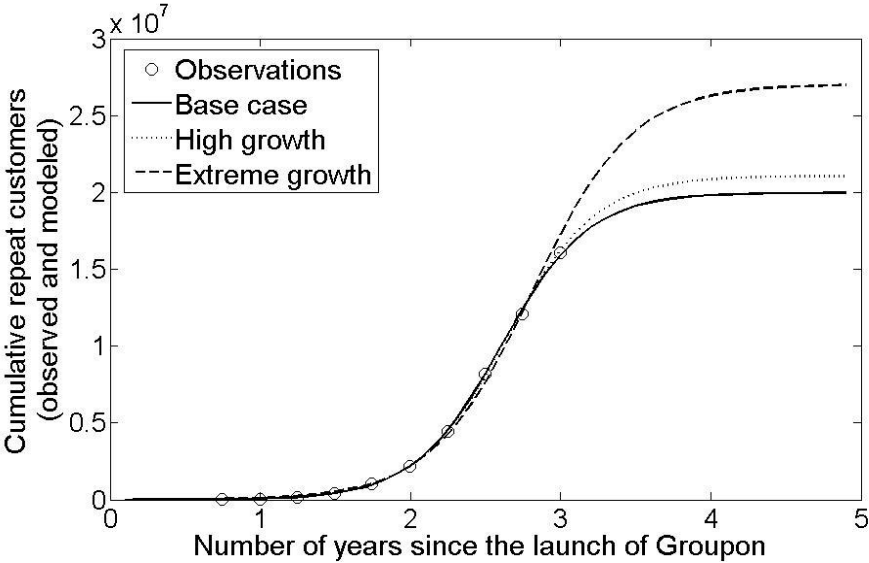

*Note: The data are taken from the SEC filing form S1, amendment 6 (October 21, 2011, page 60 from SEC [2011]). In this document, the cumulative number of repeat customers is defined as the total number of unique customers who have purchased more than one Groupon from January 1, 2009 through the end of the applicable period.*

Exhibit 11 shows the result of the analysis. It can be seen that the cumulative number of repeat customers (which is defined in Exhibit 11) follows, as in the case of facebook, a logistic function. In the base case, the plateau is reached at 17.4 million, in the high growth scenario at 21.1 million and in the extreme growth scenario at 27.0 million. At the end of the third quarter of 2011, the total number of cumulative repeat customers stood at 16.0 million. From the financial data that is available in the SEC filing report, it can be seen that there is a linear relationship between Groupon's yearly revenues and the number of cumulative repeat customers. This is clearly demonstrated in Exhibit 12. The yearly revenues per repeat customer amount to 78 USD. This means that after the completion of the growth process, when the customer ceiling is reached, the yearly revenues in the base case scenario will reach 1.4 billion USD, in the high growth scenario 1.6 billion USD and in the extreme growth scenario 2.1 billion USD.

**Exhibit 12**
**Yearly revenues of Groupon versus the number of cumulative repeat customers**

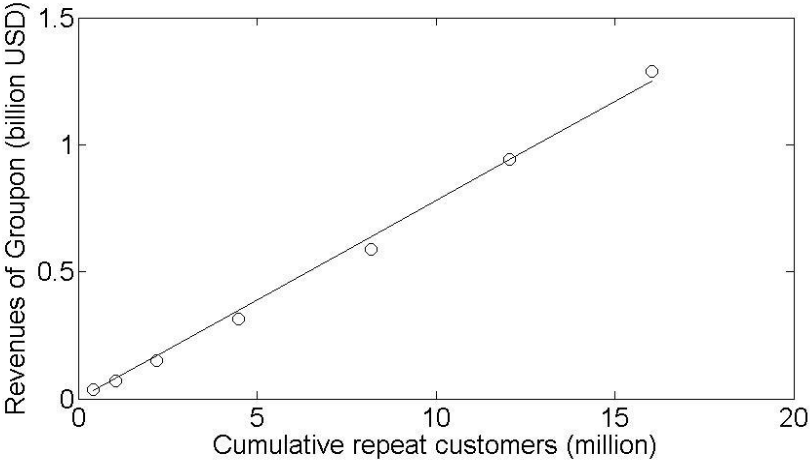

*Note: The data are taken from the SEC filing form S1, amendment 6 (October 21, 2011, page 60 from SEC [2011]). In this document, the cumulative number of repeat customers is defined as the total number of unique customers who have purchased more than one Groupon from January 1, 2009 through the end of the applicable period.*

The last step, to come to a final value, is an estimation of the profit margin. This is difficult as Groupon has mostly been reporting losses since its launch (e.g. the S1 filing report, SEC [2011], shows that for the first three quarters of 2011 total losses exceeded 200 million USD). We will use a profit margin of 20%. This is the average of all the NASDAQ listed companies at the end of October 2011. In this way, we accept that Groupon is currently investing heavily in setting up its franchise and we expect that profitability will come after this initiation period Using a discount factor of 5%, a profit margin of 20% and 78 USD of revenues per repeat customer per year gives a value of Groupon of 5.7 billion USD in the base case scenario, 6.0 billion USD in the high growth scenario and 7.7 billion USD in the extreme growth scenario.

**Conclusion**

Social networking franchises, as any other businesses, populations or new technologies face growth in competition. Such processes are characterized by an S-curve or a logistic function (Griliches [1988]; Modis [2002]). Under the constraints of competition, a limited amount of user devices, impenetrable markets and a fortiori a limited world population, the growth of their users or customers will eventually, after a period of strong initial growth, deviate from an exponential function (unlimited growth) and will follow the track of a logistic function (limited growth). This is clearly demonstrated by an analysis of the facebook users where a change of regime has occurred in 2010.

This paper proposes a new methodology to estimate the value of social networking companies taking into account the limitations to their growth. Valuations are done based on three different scenarios, a base case, a high growth and an extreme growth scenario. The methodology is demonstrated by a detailed analysis of facebook. Additionally, to demonstrate the wider applicability of the model to the whole sector, an analysis of Groupon is done. In each case, the scenarios are calibrated by fitting different logistic functions to the available user or customer data.

For facebook, we use a discount factor of 5%, a profit margin of 29% and 3.5 USD of revenues per user per year. This gives a value of 15.3 billion USD in the base case scenario, 20.2 billion USD in the high growth scenario and 32.9 billion USD in the extreme growth scenario.

These figures were chosen conservatively so as not to devalue the company unnecessarily:

- Real interest rates were put constant to 0% over the next 50 years;
- The equity risk premium stays flat at 5% over the next 50 years;
- For the revenues per user, the average of the last 5 years is taken. This disregards a clearly observed decay in revenues per user with a half life of 3.5 years;
- The profit margin remains constant at a very high 29%;
- All profit is distributed as dividend to the shareholders.

These results clearly need to be put in perspective. According to Facebook's website, a capital increase was done last January valuing the company at 50 billion USD. Rumors, spread by popular business media, value the company up to 100 billion USD. According to our model, this would imply that Facebook would need to increase its profit per user before the IPO by a factor of 3 to 6 in the base case scenario, 2.5 to 5 in high growth scenario and 1.5 to 3 in the extreme growth scenario.

It appears that facebook has already moved (at least in rhetorics) to address the issue of saturating user numbers, as founder Mark Zuckerberg has decreed recently that unique user numbers are no longer the default traffic measurement, but the volume of sharing is supposed to be a better representation of activity (PDA [2011]).

To prove the wider applicability of the methodology, we also analyzed Groupon. Using a discount factor of 5%, a profit margin of 20% and 78 USD of revenue per repeat customer per year gives a value of 5.7 billion USD in the base case scenario, 6.0 billion USD in the high growth scenario and 7.7 billion USD in the extreme growth scenario. According to its SEC filing documentation, (SEC [2011]), Groupon expects to have a market capitalization of 10.7 billion USD after the IPO. In this respect, we want to stress that a profit margin of 20% was used in our analysis even though Groupon has lost over 200 million USD in the first three quarters of 2011. Basically, at this point, the company has a -20% profit margin, which suggests that our valuations should be regarded as upper limits.

## References

Bloom, D.E., "7 Billion and counting", Science 333 (2011), pp. 562-569.

Carlson, N., "Facebook 2010 profit? Try $600 million", Business Insider (2 October 2011), http://www.msnbc.msn.com/id/41521349/ns/business-us_business/t/facebook-profit-try-million (Last visited October 5, 2011).

Dembosky, A., "Facebook puts off IPO until late 2012", Financial Times Technology (14 September 2011), http://www.ft.com/intl/cms/s/2/2b842146-dec3-11e0-a228-00144feabdc0.html?ftcamp=rss&ftcamp=crm/email/2011915/nbe/GlobalBusiness/product#axzz1Y0kJmFZs (Last visited 4 October 2011).

Facebook, "Facebook Raises $1.5 Billion", Facebook website (21 January 2011), http://www.facebook.com/press#!/press/releases.php?p=205070 (Last visited 4 October 2011).

Facebook, Facebook website (July 2011) http://www.facebook.com/press/#!/press/info.php?timeline (Last visited October 5, 2011).

Fernandez P., Aguirreamalloa J., Avendano L.C., "US Market Risk Premium Used in 2011 by Professors, Analysts and Companies: A Survey with 5731 Answers", Working Paper Series (April 8 2011), http://papers.ssrn.com/sol3/papers.cfm?abstract_id=1805852

Griliches, Z., "Hybrid Corn: An Exploration of the Economics of Technological Change", Technology, Education and Productivity: Early Papers with Notes to Subsequent Literature, (1988), New York: Basil Blackwell, pp. 27–52.

Kelly, K., "Facebook IPO Valuation Could Top $100 Billion", CNBC (13 June 2011), http://www.cnbc.com/id/43378490 (Last visited 4 October 2011).

Knowledge@Wharton, "Scrabulous and the New Social Operating System: How Facebook Gave Birth to an Industry" (23 January 2008), http://knowledge.wharton.upenn.edu/article.cfm?articleid=1883 (Last visited 4 October 2011).

Knowledge@Wharton , "Facebook's Future on the Open Market" (25 May 2011), http://knowledge.wharton.upenn.edu/article.cfm?articleid=2786 (Last visited 4 October 2011).

Mauboussin, M.J., Hiler B. "Rational exuberance? (Is there method behind the madness of Internet stock valuations?)" Equity Research, U.S./Value-Based Strategy, Credit Suisse First Boston Corporation (26 January 1999).

Modis, T., "Forecasting the Growth of Complexity and Change, Technological Forecasting & Social Change" 69(4) (2002), pp. 377-404.

Modis, T. "The Limits of Complexity and Change", the Futurist 37 (3) (2003), pp 26-32.

Modis, T. "Anticipating the Economic Turnaround with S-curves, Written for Vienna's "Die Presse" (2009), http://www.growth-dynamics.com/articles/Seifert_REVISED.pdf, (Last visited September 14. 2011)

Oreskovic., A. "Exclusive: Facebook doubles first-half revenue", Reuters (7 September 2011), http://www.reuters.com/article/2011/09/07/us-facebook-idUSTRE7863YW20110907 (Last visited 4 October 2011).

PDA, "Facebook's revenues are up, but what about user numbers?" The Guardian digital content blog (8 September 2011), http://www.guardian.co.uk/technology/pda/2011/sep/08/facebook-revenues-ipo (Last visited 4 October 2011).

Rusli, E., De La Merced M. "Groupon Plans I.P.O. With $30 Billion Valuation", NY Times Dealbook (2 June 2011), http://dealbook.nytimes.com/2011/06/02/groupon-files-to-go-public/ (Last visited October 27, 2011)

Shiller, R.J. "Irrational Exuberance", 2nd edition (2005)., Princeton University Press

SEC, Securities and Exchange Commission, Amendment No. 6 to Form s-1 (21 October 2011), Groupon Inc, http://www.sec.gov/Archives/edgar/data/1490281/000104746911008605/a2205238zs-1a.htm (Last visited October 27, 2011)

Tsotsis, A. "Report: Facebook Revenue Was $777 Million In 2009, Net Income $200 Million", TechCrunch (5 January 2011), http://techcrunch.com/2011/01/05/report-facebook-revenue-was-777-million-in-2009-net-income-200-million (Last visited October 5, 2011)

Wikipedia, "Facebook" http://en.wikipedia.org/wiki/Facebook  (Last visited October 5, 2011)